\documentstyle[revtex]{aps}

\begin{document}
\draft
\vspace{1.0cm}
\begin{center}
\Large\bf
Single - particle correlations in events with the total
disintegration of nuclei\\
\vspace{1.0cm}
\large
M. K. Suleimanov\cite{mais} , O. B. Abdinov and A. M.Bagirov\\
\large\it  Physics Institute, Azerbaijan Academy of
Sciences, Baku, Azerbaijan Republic\\
\vspace{1.0cm}
\large
A.I.Anoshin\\
\large\it
Nuclear Physics Institute, Moscow State University, Moscow, Russia\\
\vspace{1.0cm}
\large
J.Bogdanowicz\\
\large\it
Soltan Institute for Nuclear Studies, Warzaw, Poland
and Joint Institute for Nuclear Research, Dubna,Russia\\
\vspace{1.0cm}
\large
A.A.Kuznetsov\\
\large\it
Joint Institute for Nuclear Research, Dubna,Russia\\
\vspace{1.0cm}
\end{center}

\newpage
{\large

New experimental data on the  behaviour  of the  single - particle
two-dimensional correlation functions $R$ versus Q ($Q$ is the
number of nucleons emitted from nuclei) and $A_p$ ($A_p$ is the mass of
projectile nuclei) are presented in this paper .
The interactions of
protons, d, ${}^4$He and ${}^{12}$C nuclei  with carbon  nuclei
( at a momentum of 4.2 A GeV/c) are considered.  The values of $R$ are
obtained separately for $\pi^-$ mesons and protons.  In so doing,  the
values of $R$ are normalized so that $-1\leq R \leq 1$.  The value of
$R=0$ corresponds to the case of the absence of correlations.  It has
been found that the $Q$- and  $A_p$-dependence of $R$ takes
place only for weak correlations ($R< 0.3$).
In the main (90 $\%$),these correlations are connected with the
variable $p_t$ and have a nonlinear character, that is the regions with
different characters of the $Q$-dependence  of $R$ are separated:
there is a change of regimes in the $Q$-dependences of $R$.
 The correlations weaken with increasing $A_p$ , and the
variable $R$ gets the least values of all the considered ones in ${}^{12}CC$
interactions.  Simultaneously with  weakening the correlations in the
region of large $Q$ ,  the character of the Q-dependence of $R$
changes.

\noindent The work has been performed at the Laboratory of High
Energies, JINR.}
\vspace{0.5cm}

{\large\noindent PACS number(s): 25.70.Np, 25.45.-z, 22.55.-e
}

\begin{center}
{\Large\bf I. INTRODUCTION}
\end{center}
{\large

The aim of this paper is to give information on
the values of  single-particle correlation functions ($R$) in
nucleus-nucleus collisions and in events with TDN .

The study of
correlation effects in particle production is the source of
information on the dynamics of interactions supplementing the
information obtained from the single-particle inclusive distributions,
from the analysis of the mean characteristics and so on. [1].
\newpage
\noindent In this paper, we present the results of investigating the
Q-dependence \footnotemark \footnotetext{\large The values of $Q$ were
determined as $Q$ is $Q=N_{+}-N_{\pi^-}$ ( $N_{+}$ and $N_{\pi^-}$ are
the numbers of positive charged particles and  $\pi^-$ mesons ,
respectively ) .The variable $Q$ is proportional to the number of
protons emitted from nuclei during the interaction.
In detail, using the variable $Q$ ,see paper [2] .} of $R$ for
$\pi^-$ mesons and protons emitted in nucleus-nucleus collisions.
Such an experiment allows us to obtain  necessary information.

Interest in the processes of TDN is determined by that they are extreme cases.
It is supposed that in these cases the anomalously large
densities of nuclear  matter can be realized, and the effects associated
with  the collective properties of nuclear matter are revealed. We think
that the processes of TDN correspond to qualitatively new
states of nuclear matter, and the transition to these states goes
through the "critical" values of  $Q\rightarrow Q^{*}$.  The presence
of $Q^{*}$ should lead to changing in the $Q$ dependences of event
characteristics in the region near - $Q^{*}$. Therefore, it is offered
to use the condition $Q\geq Q^{*}$ as an event selection criterion with
the total disintegration of  target nuclei.

  The probability distributions of observing
pC, dC, ${}^4$HeC and ${}^{12}$CC events with the given
values of the variable $Q$
are studied in papers [2].
The dependences
of the mean characteristics and the inclusive spectra of secondary $\pi$
mesons and protons on $Q$  are studied in papers [3-4]
\footnotemark \footnotetext{\large For this, the groups of events with

\begin{center}
$Q\geq 1 ;2 ; 3 ;...$
\end{center}

\noindent were defined. Then,
the  mean characteristics and the invariant inclusive spectra were
obtained for each group.}.
The  results , obtained in these papers , confirm our
assumption that a  critical value of $Q^*$ for $Q$ exists (its
excess leads to the TDN ) .

  The single-particle two-dimensional correlation function $R(~x,z~)$
is used in the analysis. It is connected with  that  the correlations
strengthen when passing to high orders of $R$ [5-7]. The values of
$R(~x,z~)$  for the variables $x,z$ were calculated by the following
formula:

\begin{center}
$R(x,z) = (Exz - ExEz)/ (\sigma(x) \sigma(z))$
\end{center}

Here   $Exz$ is the mixed  mathematical expectation of quantities $x$
and $z$ ; $Ex$  and  $Ez$ are the mathematical expectations of $x$ and
$z$, respectively ;  $\sigma(x)$ and $\sigma(z)$ are the r.m.s. deviations
of $x$ and $z$ , respectively.

To evaluate $Exz, Ex, Ez$, $\sigma(x)$ and $\sigma(z)$,
the following formulae were used :

\begin{center}
$Exz = (1/N) \sum_{i=1}^{40} \sum_{j=1}^{40} N_{ij} x_i z_j$
\end{center}

\begin{center}
$Ex = (1/N) \sum_{i=1}^{40} N_i x_i$
\end{center}

\begin{center}
$Ez = (1/N) \sum_{j=1}^{40} N_{j} z_j$
\end{center}

\begin{center}
$\sigma(x) = (Ex^2 - (Ex)^2)^{1/2}, \sigma(z) = (Ez^2 - (Ez)^2)^{1/2}$
\end{center}

\begin{center}
$Ex^2 = (1/N) \sum_{i=1}^{40} N_{i} x_i^2,$
$Ez^2 = (1/N) \sum_{j=1}^{40} N_{j} z_j^2$
\end{center}

In these formulae , $N_{ij}$ is the number of particles  hitting an
$(i,j)$ -th cell .

\begin{center}
$N = \sum_{i=1}^{40} \sum_{j=1}^{40} N_{ij},$
$N_i = \sum_{j=1}^{40} N_{ij},$
$N_j = \sum_{i=1}^{40} N_{ij}$
\end{center}

During  the  construction of the two-dimensional distributions for the
chosen variables , the intervals , corresponding to them were divided
into  40 subintervals.}

\vspace{1.0cm}

\begin{center}
{\Large\bf II. METHODS OF THE EXPERIMENT}
\end{center}
\vspace{1.0cm}

{\large
The experimental data have been obtained from the 2-m propane bubble
chamber of  LHE, JINR . The chamber placed in a magnetic field of
1.5 T, was exposed to beams of light relativistic nuclei at the Dubna
Synchrophasotron.
Practically all secondaries emitted at a 4$\pi$ total solid angle were
detected in the chamber. All negative particles, except identified
electrons, were considered as $\pi^-$ mesons. The contaminations by
misidentified electrons and negative strange particles do not exceed
5$\%$ and 1$\%$, respectively.The average minimum momentum for pion
registration is about 70 Mev/c. The protons were selected by the
statistical method applied to all positive particles with a momentum
of $p>$500 MeV/c ( we identified slow protons  with
$p\le$700 Mev/c by ionization in the chamber).
In this experiment, we used  5284   $pC$ , 6735   $dC$ , 4852  ${}^4HeC$
and 7327
${}^{12}CC$ interactions at a momentum of 4.2 A GeV/c ( for methodical
details see [9]) .  The available statistical material was separated
into  groups of events with the following values of Q:

\begin{center}
$Q\geq 1 ;2 ; 3 ;..Q^*;...$       (1)
\end{center}

The values of $R$ were determined for $\pi^-$  mesons and protons
from these groups of events.  In this case , we considered
only $\pi^-$  mesons and protons with the errors in measuring the
momenta not exceeding 30 $\%$.

 We considered the following correlation functions
$R(~p,\theta~)$, $R(~p,p_t~)$ ,
$R(~p,y~)$ , $R(~p,\beta~)$, $R(~\theta,p_t~)$ , $R(~\theta,y~)$, $R(~\theta,\beta^0$~),
$R(~p_t,y~)$   ,$R(~p_t,\beta^0~)$ ,$R(~y,\beta^0~)$ , where :

   -  $p$ are  momenta in the laboratory coordinate system (lcs)

   - $\theta$   -   emitted angles in the lcs

   - $p_t$      -  transverse momenta,

   - $y$        -   rapidities  in the lcs.

   $\beta^0$ - orders of cumulativity ( here $\beta^0$ = (E-$p_L$) /
$m_N$ , $E$ is the total energy (in the lcs),  $p_L$  is the longitudinal
momentum (in the lcs) and $m_N$  is the nucleon mass ).

  The correlations between $p_t$ and $p_L$  for  $\pi^-$ mesons ,
produced in  $dC$  interactions at 1.7 and 4.2 GeV/c per nucleon , were
  studied in paper [10]. The dependence of the
distribution density $\pi^-$ mesons on  these variables was  observed.
The forms of these distributions turned out to depend on beam
energy.   The  $y$ and
 $p_t$ distributions of protons were measured in  S+W, O+W
  and p+W reactions at 200 GeV/A  [11]. The density of
    the $y$ distribution was found to grow linearly  with increasing
 transverse energy for  all the 3 reactions . However, the slope in
  p+W is  sharper  than in O +W and S+W.  The rapidity density  in
  p+W is much larger than it was predicted on the basis of summing
    nucleus-nucleus collisions without taking into account
  nuclear effects pointing to the importance of
  rescattering effects.  The  results obtained in papers [10 - 11]
  show a good perspective of using the $R$ function for
  revealing  qualitatively new phenomena in interactions of
  relativistic nuclei.

In our experiment ,  the parameters
  $\beta^0, \theta, y$ were chosen in the intervals :

$0\leq \beta^0 \leq 3; 0\leq \theta \leq 180^0$ and $ -2 \leq y \leq
3.5$

in all cases.

The parameters $p$ and $p_t$ in $pC , dC, {}^{4}HeC, {}^{12}CC$
interactions were chosen from the intervals :

$0.07 GeV/c \leq p \leq 10 GeV/c$ and $ 0.07 GeV/c \leq p_t \leq 4 GeV/c$

\noindent both for protons and for $\pi^-$ mesons.

\vspace{1.0cm}

\begin{center}
{\Large\bf III. THE RESULTS OF THE EXPERIMENT AND DISCUSSION}
\end{center}
\vspace{1.0cm}

{\large
 The values of $R$ at  different $Q$ for  different pairs were obtained
using the method described above. In all ,  the
$Q$-dependences for 100 $R$ functions were obtained (  10 types of
$R$ functions separately for $\pi^-$  mesons and protons in 5 types of
interactions, see above). According to the character of the  $Q$ - dependence
of $R$ , the data can be divided into two
groups : group I - the data on the Q independence of $R$   and group II -
the data showing  the $Q$ dependence of $R$ . The data from the first
group were fitted with an $a*Q + b$ expression , where $a$ ( for
this group of data , the values of $a$ turned out to
be close to
\newpage
\noindent zero.) and $b$ are the fitting parameters
\footnotemark \footnotetext{\large We do not
present the measured  errors  of $b$ but take them
into account in our  conclusions.}.

The values of $b$ are presented
in tables I and II.  The cases from the second
group are denoted by dashes.  The absolute values of $R$
depending on $Q$ for this group of data are shown in figures
1-5 ( the curves are  hand-drawn )
\footnotemark \footnotetext{\large
We do not present  the  data on
a) the values of $R(p,\beta^0)$ depending on
$Q$ in $pC$ interactions (for $\pi^- $ mesons and protons) and $dC$
interactions (for protons). For these cases , the values of
$R(p,\beta^0)$ decrease with increasing $Q$. In so doing , this
dependence has a linear character for $\pi^-$ mesons  and almost a
logarithmic character for protons ; b) the values of $R(p_t,\beta^0)$ as
in this case the behaviour of $R$ versus  $Q$ for  $pC$ , $dC$ and
${}^4HeC$ interactions is similar to those shown in figs.  3-5.  For
${}^{12}CC$  interactions , the behaviour of $R(p_t,\beta^0)$ as a
function of  $Q$ is a line with "break" in its character.  }.

From the data in tables I-II, one can  see the following:

       -  the $Q$-dependence of $R$ is not
observed in 71 $\%$ cases (of 100, group I of data), and
 the function $R$ respectively depends on $Q$ (group II of data)
in 29 $\%$ cases ,

       -  in 90 $\%$ cases of group II the $Q$-dependence of $R$ is
mainly due to  the variable $p_t$,

       -  the data from  group I also point  to the independence of the
behaviour of projectile  mass  ($A_p$)  in nucleus-nucleus
interactions.

  From figs. 1-5, one can draw  the following
conclusions :

      -  in 75  $\%$ cases , the $Q$-dependence of $R$ has a nonlinear
character , i.e. the regions with  different $Q$-dependences of $R$
 are separated  or the change of regimes takes place in these
dependences.  The totality of these data allows one to determine the
" critical"  values of $Q=Q^*$  corresponding to  the transition
 from one region to another.
These
values of $Q^*$  mainly coincide with those  obtained in previous papers
 [1-3] and are  used for event selection with TDN.
Thus , we  have that the
correlation analysis also confirms that  events with  TDN qualitatively
 differ from "usual" events, and  it is necessary to use  condition (1)
for their separation.
This confirms
our main affirmation that the TDN processes are those in which  so a
large ("critical" ) fraction  of nuclear nucleons is emitted whose excess
leads to showing  qualitatively new properties (see paper [1]). In
particular , from the analysis results presented in the above figures ,
it has been found that

      -  82 $\%$ of cases from  group II have   $R < 0.3 $,
i.e.  weak correlations  related to  the variable
$p_t$ and  depending on $Q$ , mainly take place ;

       - in all the considered cases , a strong change of the
form of the $Q$- and $A-p$-dependences of  $\mid R \mid$  is observed.
For example ,
the correlations weaken with increasing $A_p$ , and the variable $R$
gets the  least values of all the considered ones in ${}^{12}CC$ interactions
( this result agrees well with the  conclusions of paper [11]).  The
character of the $Q$ - dependence of $\mid R \mid $ also changes
simultaneously with weakening the correlations in the region of large
$Q$ ( TDN region ) This dependence is a  line with "break" for $pC$ ,
$dC$ interactions , it is of the step-by step form for ${}^4HeC$
interactions and the "zigzag" form for ${}^{12}CC$ interactions.  It is
possible that the "zigzag" form is the result of the influence of
density fluctuations of nuclear matter in the TDN region  on these
dependences . In earlier paper [12]  (in the studies of the
multiplicity distributions and their second moments for negative charged
particles produced in "central" collisions ("central") and in
interactions of minimum trigger in ${}^{32}S+S $ collisions at 200 A GeV
over different rapidity intervals ), it has been found that the models
of FRITIOF and VENUS mainly  describe the dependence of second moments
on rapidity intervals for the events with minimium trigger and not  for
"central". The conclusion has been drawn that the behaviour of second
moments for "central" indicates  increasing the multiplicity
fluctuation . These observations support the conclusions from the
analysis of entropy.  The entropy for central ${}^{32}S+S $ is larger
than that expected in the models.  The results of the present paper also
confirm this conclusion.  We also think that the "zigzag" form in events
with the total disintegration of nuclei can be connected with the
density fluctuations  of nuclear matter at large $Q$.

Under these experimental conditions , we observed
a strong $Q$-dependence of the mean values of the kinetic energy of
$\pi^-$ mesons in ${}^{12}CC$ interactions
in the region $Q\ge Q^*$ - of the total disintegration of nuclei.
The present results show  that this is possibly due to
the fluctuations of nuclear density in events with the total
disintegration of nuclei.  It has already been concluded  [13] that the
transparency of nuclear matter in "central" decreases significantly. In the
authors' opinion , it testifies   of a high baryon density  reached
in the  investigated interactions.  We assume that at our energies
they are mixed  states  corresponding to different degrees
of freedom ,  as well as  quark - gluon  degrees of freedom;

       -  for protons, the behaviours of $R(\theta,p_t)$ (fig.
4) and $R(p_t,y)$ (fig. 5) versus  $Q$  are similar and
differ from the behaviour of the $Q$ - dependence of $R(p,p_t)$ ( fig.
3).}
\vspace{1.0cm}

\begin{center}
{\Large\bf ACKNOWLEDGEMENTS}
\end{center}
\vspace{1.0cm}

{\large
    The authors consider it their  pleasant debt  to thank the
staff of the 2-m propane bubble chamber  for giving the experimental
material, N.  Amelin and A. Golokhvastov for useful discussions ,
O. Rogochevsky  for his help and also Academician A. M. Baldin for
his permanent attention to work.}

\vspace{1.0cm}

\begin{center}
\line(1,0){250.}
\end{center}
\vspace{1.0cm}

\begin{itemize}

\item [[ 1]] E. A. De Wolf , I. M. Dremin , W. Kittel .
Russian Journal UFN , {\bf 163} ,  3 , (1993)
\item [[ 2]]  O. B. Abdinov. {\it et al.} , JINR Rapid
Communications  No. {\bf 1[75]} , 51 , (1996) .
\item [[ 3]]  O. B. Abdinov {\it et al.} ,
JINR Rapid Communications  No. {\bf 1[81]} ,109 , (1997) .
\item [[ 4]]  O. B.
Abdinov  {\it et al.} , JINR Communications No. {\bf E1-97-178} , 1,
(1997) .
\item [[ 5]] P. Abreu {\it
et al.} , Phys.  Lett. B{\bf 247} , 137 , (1990) ;
A. De Angelis  Mod. Phys. Lett. A{\bf 5} , 2395 , (1990).
\item [[ 6]] I. V. Ajienko {\it et al.} , Phys. Lett. B{\bf 222} ,
 306 , (1989) ; I. V. Ajienko {\it et al.} , Phys. Lett. B{\bf235} , 373
 , (1990) ;  N. M. Agababyan {\it et al.} , Phys. Lett. B{\bf 261} ,
 165 , (1991).
\item [[ 7]] C. Albajar {\it et al.} , Nucl. Phys. B{\bf 345} , 1 , (1990).
\item [[ 8]] The BBCDHSSTTU-BW Collaboration , A. U.
 Abdurakhimov {\it et al.} , Phys. Lett. , B{\bf 39} , 371 , (1972).
\item [[ 9]] N. Akhababian {\it et al.} , JINR Preprint {\bf
1-12114} , (1979).
\item [[ 10]] K. N.  Ermakov  , O. V.
 Rogochevskiy , M. V.  Stabnikov , Preprint RAS , {\bf 2012} , 1
 ,(1994).
\item [[ 11]] T.  Akesson  {\it et al.} , Z.  Phys.  C. {\bf
53} , 183 , (1992).
\item [[ 12]] J. Bachler {\it et al.} , Z.  Phys.
C.  , {\bf 56} , 347 , (1992)
\item [[ 13]] J.  Berrette {\it et al.} , Phys.Rev. C.  {\bf 50} , 3047
, (1994).

\end{itemize}

\newpage

\figure{$Q$-dependence of  $\mid R(\theta,p_t)\mid$
for $\pi^-$ mesons in $pC$  , $dC$  ,${}^4HeC$ and ${}^{12}CC$ interactions.
The values of R shown at $Q=1;2;3;...$ correspond to the groups
of events with $Q\ge1;2;3;...$ , respectively.
\label{fig1}}

\figure{ $Q$-dependence of  $\mid R(p_t,y)\mid$ for $\pi^-$ mesons
in $pC$  , $dC$  ,${}^4HeC$ and ${}^{12}CC$ interactions.
The values of R shown at $Q=1;2;3;...$ correspond to the groups
of events with $Q\ge1;2;3;...$ , respectively.
\label{fig2}}

\figure{$Q$-dependence of $\mid R(p,p_t)\mid$ for protons
in $pC$  , $dC$  ,${}^4HeC$ and ${}^{12}CC$ interactions.
The values of R shown at $Q=1;2;3;...$ correspond to the groups
of events with $Q\ge1;2;3;...$ , respectively.
\label{fig3}}

\figure{$Q$-dependence of $\mid R(\theta,p_t)\mid$ for protons
in $pC$  , $dC$  ,${}^4HeC$ and ${}^{12}CC$ interactions.
The values of R shown at $Q=1;2;3;...$ correspond to the groups
of events with $Q\ge1;2;3;...$ , respectively.
\label{fig4}}

\figure{$Q$-dependence of $\mid R(p_t,y)\mid$ for protons
in $pC$  , $dC$  ,${}^4HeC$ and ${}^{12}CC$ interactions.
The values of R shown at $Q=1;2;3;...$ correspond to the groups
of events with $Q\ge1;2;3;...$ , respectively.
\label{fig5}}

\newpage

\begin{center}
{\Large  TABLE I.  For $\pi^-$ mesons}
\end{center}

\begin{center}

\line(1,0){177.}
\vspace{0.1cm}

\begin{tabular}{ccccc} \hline
Type   & \multicolumn{1}{c}{$pC$} & \multicolumn{1}{c}{$dC$}      &  \multicolumn{1}{c}{${}^4HeC$} &\multicolumn{1}{c}{${}^{12}CC$} \\
of events. & & & & \\ \hline
$R(p,\theta)$  &     - 0.49 &  - 0.52 &    - 0.49 &     - 0.48 \\
$R(p,p_t)$          &      0.62 &     0.52 & 0.49 &      0.50 \\
$R(p,y)$            & 0.67 &      0.71 & 0.70 & 0.69      \\
$ R(p,\beta^0)$ &         -    &     - &    -0.22 &     -0.19   \\
$R(\theta,p_t)$     &      - &     -    &    -    &       -      \\
$R(\theta,y)$       &    -0.93  &    -0.92  &  -0.91  &     -0.90   \\
$R(\theta,\beta^0)$ &  0.83   &     0.78   &  0.79   &     0.78    \\
$ R(p_t,y)$ &           -    &       -    &     -    &     -      \\
$R(p_t,\beta^0)$    &   0.31  &      0.42  &   0.43  &     0.45    \\
 $R(y,\beta^0)$     &  -0.79  &     -0.73  &   -0.73 &     -0.71  \\
\hline
\end{tabular}

\vspace{0.1cm}
\line(1,0){177.}

\end{center}

\vspace{1.0cm}
\begin{center}
{\Large  TABLE II. For protons}
\end{center}

\begin{center}

\line(1,0){166.}
\vspace{0.1cm}

\begin{tabular}{ccccc} \hline
Type    & \multicolumn{1}{c}{$pC$} & \multicolumn{1}{c}{$dC$}      &  \multicolumn{1}{c}{${}^4HeC$} &\multicolumn{1}{c}{${}^{12}CC$} \\
of events  & & & & \\ \hline
$R(p,\theta)$   &     -0.67  &  -0.68  &     -0.63  &   -0.63              \\
$R(p,p_t)$      &     -    &        -    &      -    &      -                   \\
$R(p,y)$        &    0.97  &       0.96  &    0.93  &      0.93                \\
$ R(p,\beta^0)$  &      -    &     -0.85  &      -0.80  &   -0.78               \\
$R(\theta,p_t)$  &         -  &         -  &         -  &      -                   \\
$R(\theta,y)$    &       -0.80  &    -0.81  &       -0.80  &   -0.82            \\
$R(\theta,\beta^0)$  &       0.90  &      0.90  &     0.92  &     0.93              \\
$ R(p_t,y)$       &    -   &         -   &         -   &    -                  \\
$R(p_t,\beta^0)$  &    -    &        -    &       -    &   -                  \\
 $R(y,\beta^0)$   &    -0.94 &      -0.95 &     -0.94  &     -0.94              \\
\hline
\end{tabular}
\vspace{0.1cm}
\line(1,0){166.}
\end{center}
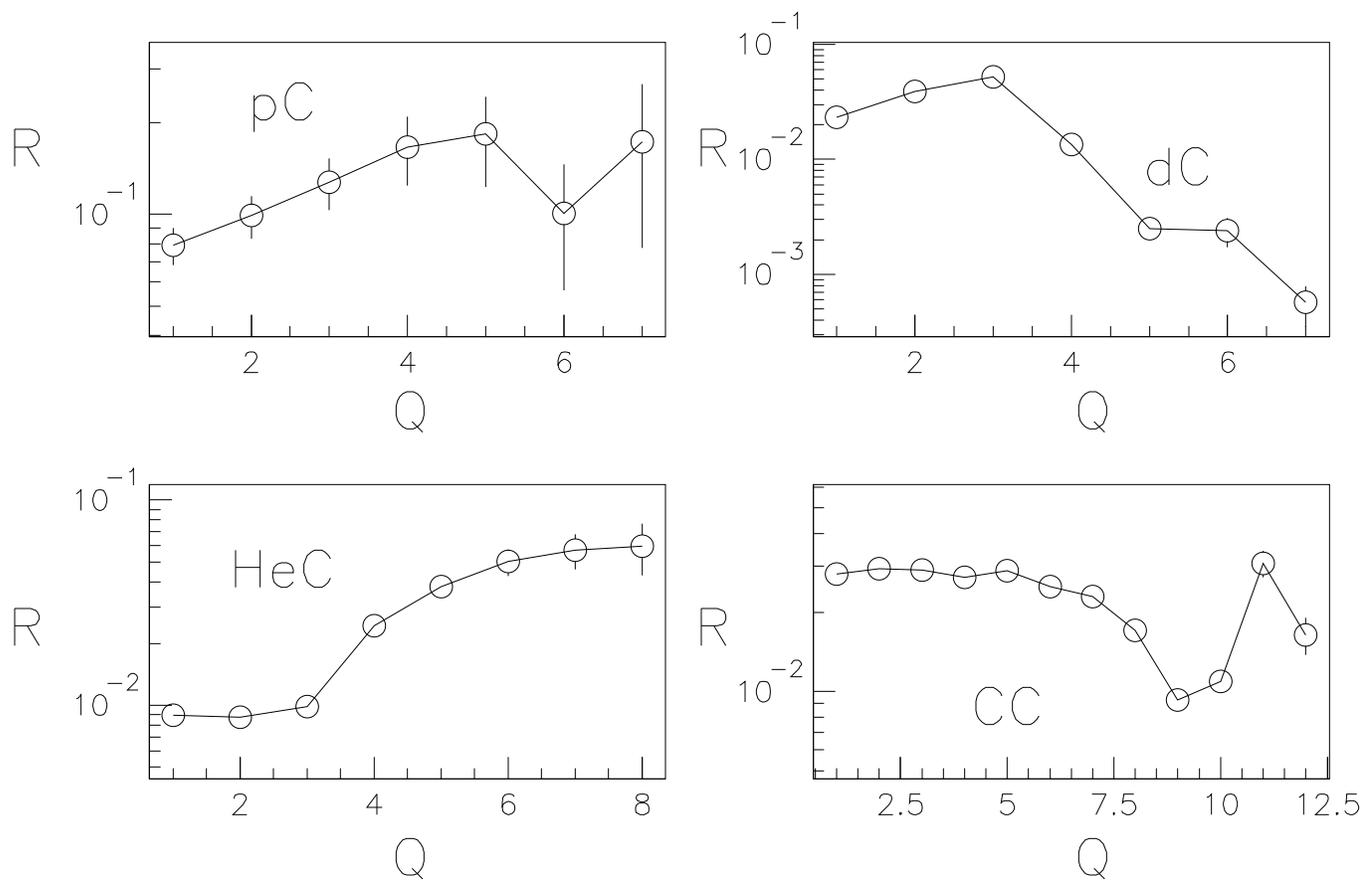
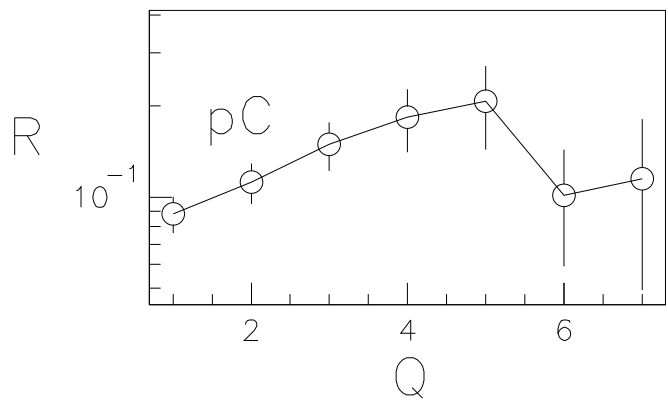
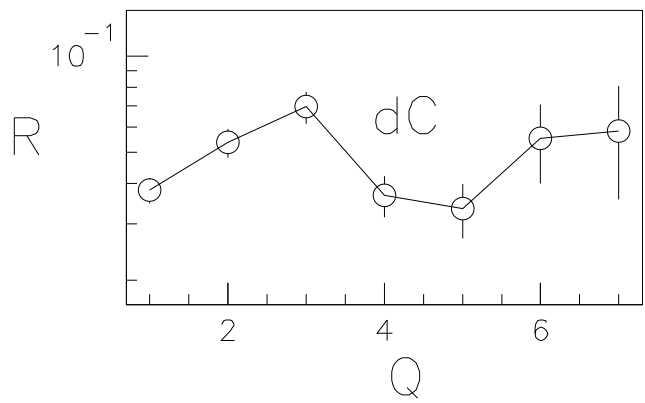
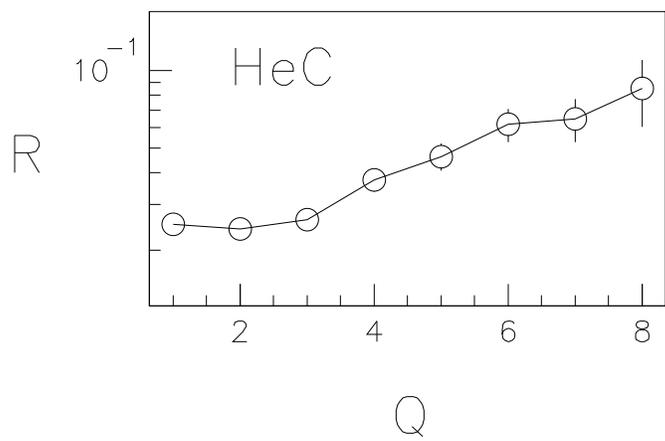
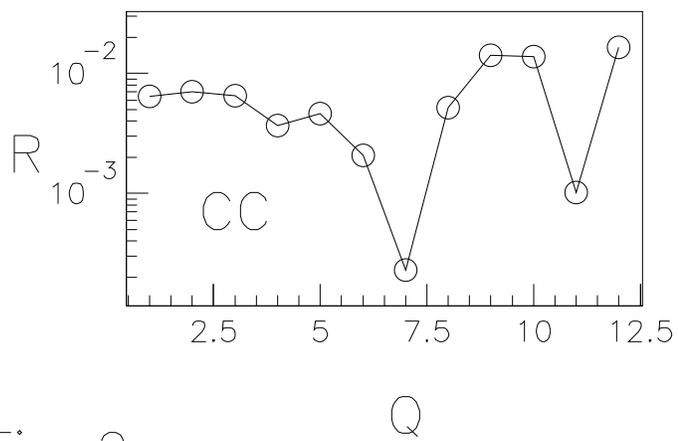
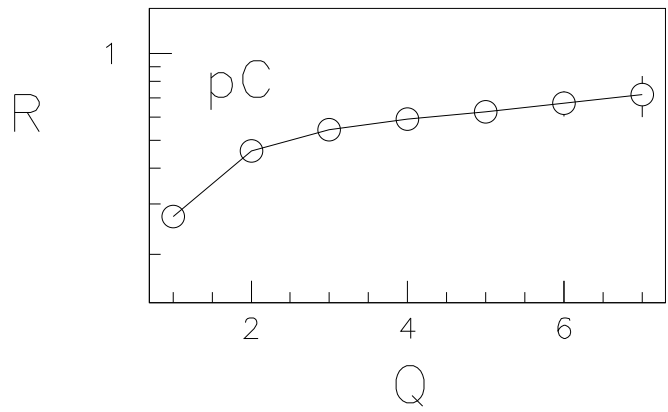
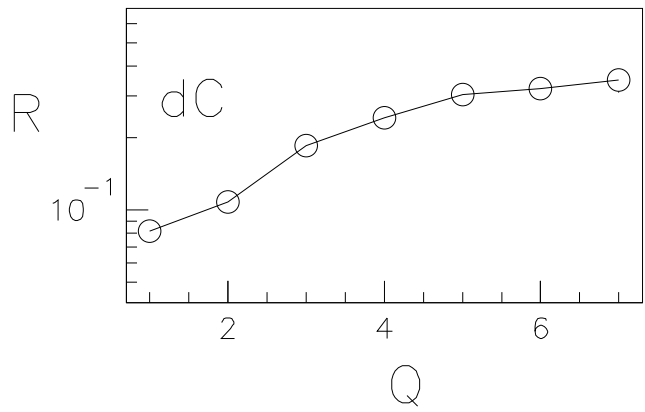
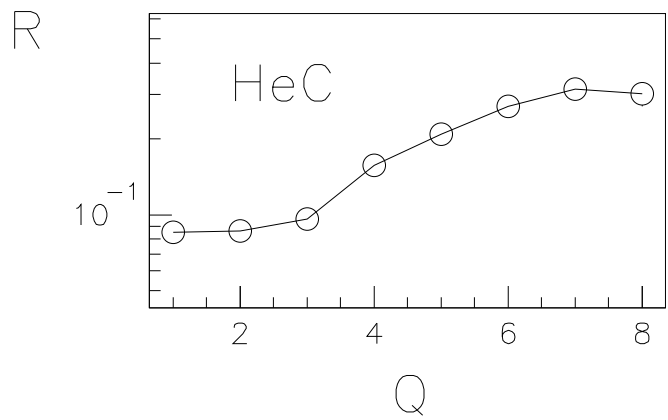
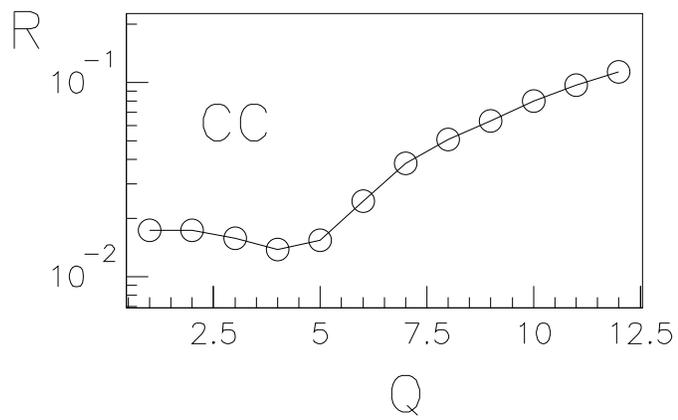
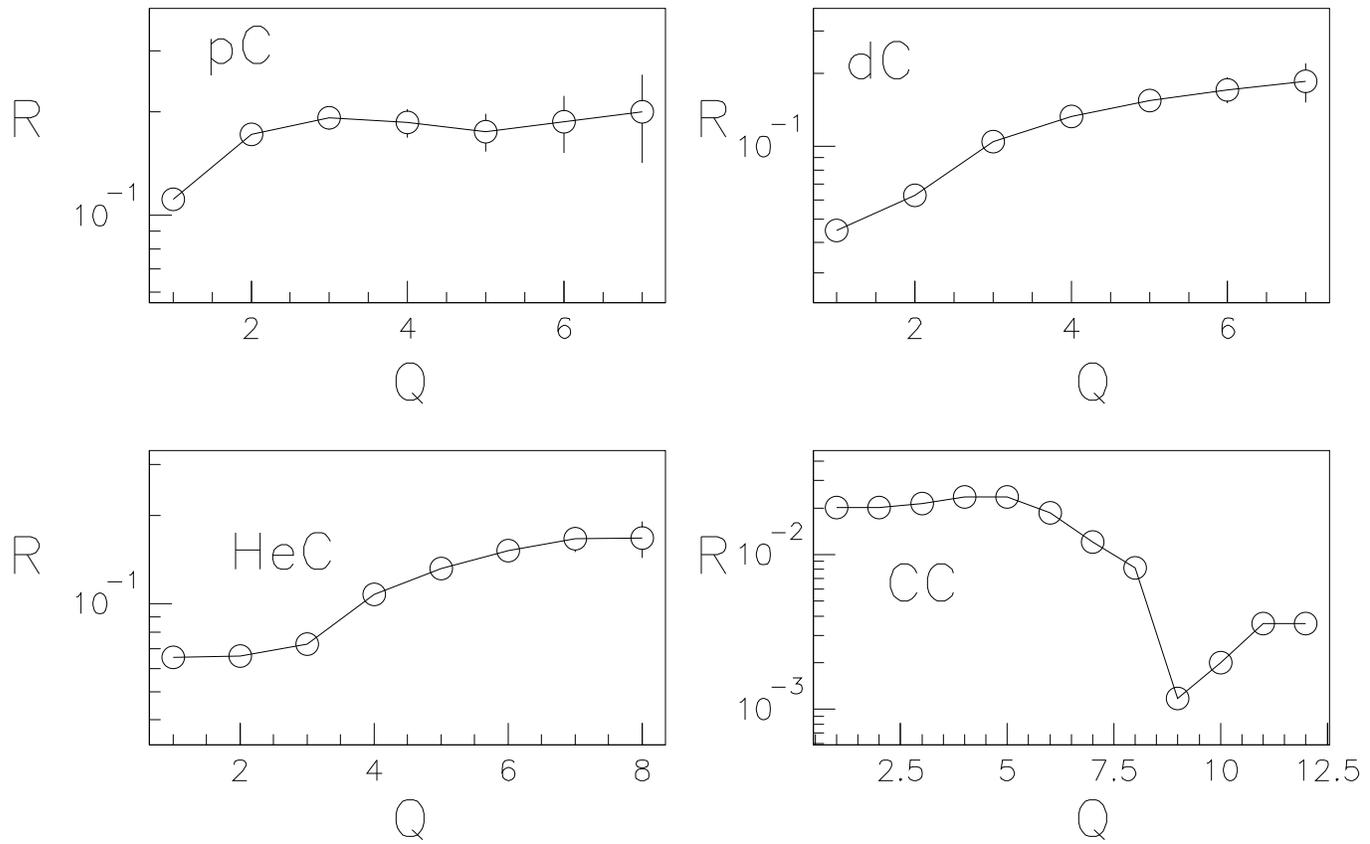
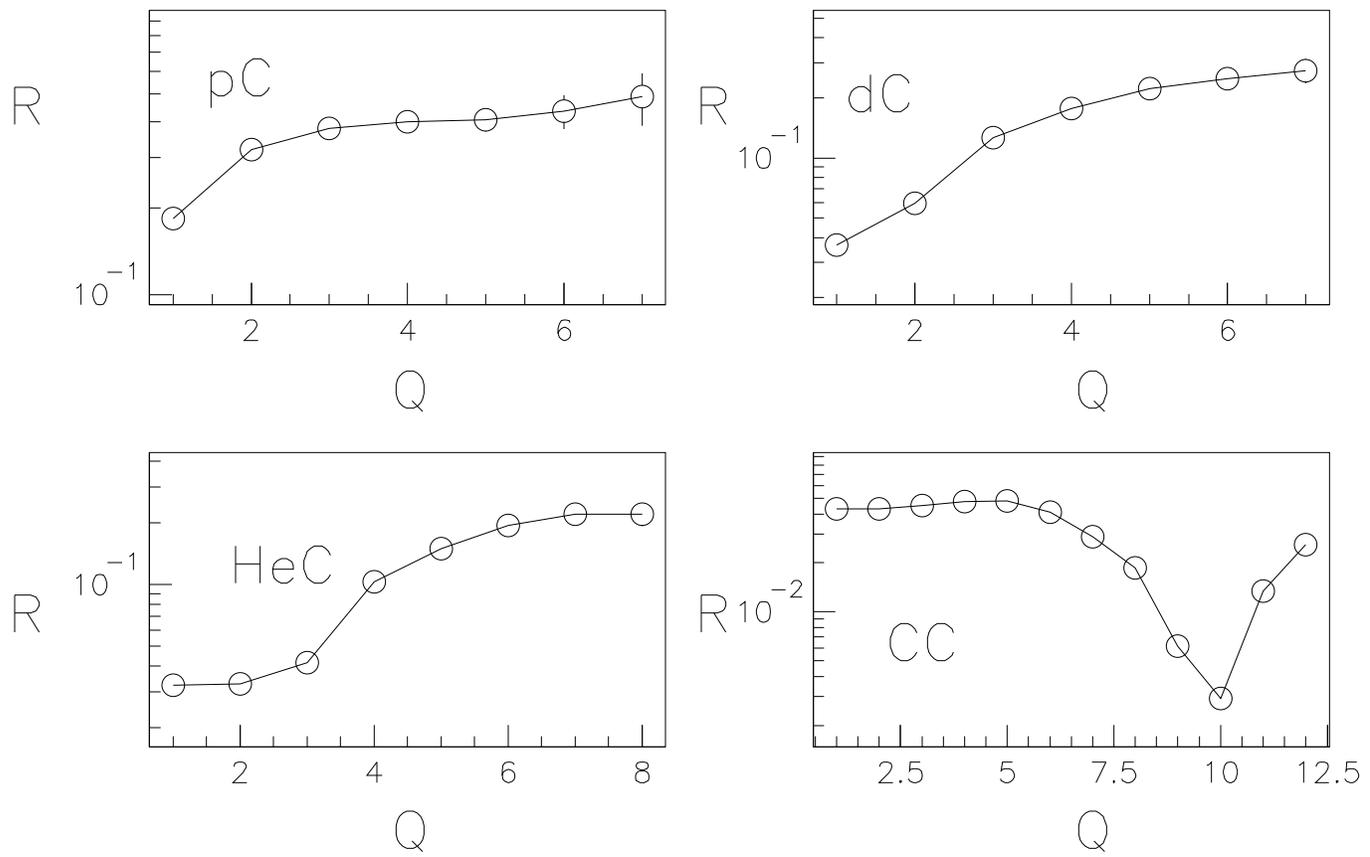
\end{document}